\documentclass[
	fontsize=12pt,
	oneside,
	DIV=classic,
	paper=a4,
	pagesize=auto,
	titlepage=yes,
	bibliography		= totocnumbered,
	listof					= totocnumbered,
	parskip					= half,
	abstract				= yes
	]{scrartcl}

	\newcommand{\cpfversion}{Version 0.9}
	\newcommand{\cpfbuild}{20231229}

	\usepackage[utf8]{inputenc}
	\usepackage[T1]{fontenc}
	\usepackage[english]{babel}
	
	\usepackage[figure,table]{totalcount}

	\usepackage{times}

	\usepackage[activate={true,nocompatibility},final,tracking=false,kerning=true,spacing=true,factor=1100,stretch=10,shrink=10]{microtype}
	
	\usepackage{enumerate}
				
	\usepackage{csquotes}

	\usepackage[nottoc]{tocbibind}

	\usepackage{hyperref}

	\usepackage[headsepline]{scrlayer-scrpage}
	\ohead{Fries, Christian P.}
	\ihead{Implied CO$_{\textbf{2}}$-Price and Interest Rate of Carbon}
	\ifoot{\tiny \copyright 2023 Christian Fries}
	\ofoot{\tiny \cpfversion\ (\cpfbuild )\\ \href{http://www.christianfries.com/finmath}{http://www.christianfries.com/finmath}}

	\usepackage[backend=bibtex]{biblatex}   
	\bibliography{literature}

	\usepackage[intlimits]{amsmath}
	\usepackage{amssymb}

	\usepackage{algorithmic,algorithm,listings}
	
	\usepackage{graphicx,xcolor,subfig}
	
	\definecolor{DarkBlue}{rgb}{0.0, 0.0, 0.5}
	\definecolor{DarkRed}{rgb}{0.5, 0.0, 0.0}
	\definecolor{DarkGreen}{rgb}{0.0, 0.5, 0.0}
	\definecolor{DarkYellow}{rgb}{0.5, 0.5, 0.0}
	\definecolor{Brown}{cmyk}{0.00,0.80,1.00,0.60}
	\definecolor{DarkGreen}{cmyk}{0.64,0.00,0.95,0.60}
	\definecolor{DarkBlue}{cmyk}{0.70,0.60,0.00,0.60}

	\usepackage{ifpdf}	
						
	\ifpdf
		\usepackage{hyperref}
		\hypersetup{
			hyperindex = true,
			pdfpagelayout=TwoPageRight,
			pdfborderstyle={/S/U/W 1},
			colorlinks = true,
			linktocpage = true, linkcolor=blue, citecolor=blue, urlcolor=blue, bookmarks,
			pdfauthor={Christian Fries},
			pdfsubject={A Short Note on the Social Cost of Carbon},
			pdftitle={Implied CO2-Price and Interest Rate of Carbon},
			pdfstartview=FitH
		}
	\else
		\usepackage{url}
	\fi

	\newcounter{cpf_counter} 
	\newcounter{cpfNumberOfFigures} 
	\newcounter{cpfNumberOfTables} 

\usepackage{tikz}
\usetikzlibrary{positioning,arrows,shapes,calc,matrix}
\usetikzlibrary{shapes.misc,math}
\usepackage{pgf-umlcd}
\usepackage{pgfplots}
\pgfmathdeclarefunction{invgauss}{2}{%
  \pgfmathparse{sqrt(-2*ln(#1))*cos(deg(2*pi*#2))}%
}
\usepackage{adjustbox}

\begin{document}
	
		\subject{The Social Cost of Carbon \\ does not cover the social cost of carbon}
		\author{
			Christian Fries
			\thanks{\url{http://www.christianfries.com}}
			\thanks{Department of Mathematics, University of Munich}
		}
		\title{Implied CO$_{\textbf{2}}$-Price\\[1.2ex] and\\[1.2ex] Interest Rate of Carbon}
		\subtitle{
			 \centerline{\small \cpfversion}
		}
		\date{December 22, 2023 \\ (This version: January 27, 2024)}

	\maketitle

	\begin{abstract}
		By its nature, the so-called social cost of carbon (SCC(t)) will likely not cover the cost induced by climate change (damage cost and abatement cost) if it is used as a CO$_2$-price. It is a marginal price only.

		We define an implied CO$_2$-price that covers the climate change-induced costs. The price can be interpreted as a \textit{polluter pays principle}.

		A numerical analysis using a classical DICE model reveals that the cost-implied CO$_2$ price is around 500 \$/tCO$_2$, while the corresponding price associated with the SCC is about 50 \$/tCO$_2$.

		In addition, we define the internal rate of return of carbon abatement and calculate it for the classical DICE model. This rate is much higher than the model's discount rate, which may suggest the advantage of financing abatement by loans.
	\end{abstract}

	%
	\microtypesetup{protrusion=false}
	\tableofcontents
	\microtypesetup{protrusion=true}

	\clearpage
	\section{Introduction}
	
	The \textit{Social Cost of Carbon} (SCC) \cite{Nordhaus2017,RennertErricksonPrest2022} derived from an integrated assessment model (IAM) (like the DICE Model) is defined as
	\begin{equation*}
		\frac{\partial V(0)}{\partial E(t)} \big/ \frac{\partial V(0)}{\partial Z(t)} \text{,}
	\end{equation*}
	where $V(0)$ is the total discounted welfare, $E(t)$ is the time $t$ emissions measured in tCO$_{\text{2}}$ and $Z(t)$ is the time $t$ consumption measured in Dollar.\footnote{Note that we use the letter $Z$ for the consumption, while \cite{Nordhaus2017} uses the letter $C$ for consumption. We made the change as we will denote the total cost (damage and abatement costs) by $C(t)$. Consumption will not play a role in this paper - except for stating this formula here once.}

	\smallskip

	The SCC is sometimes considered a good choice for a CO$_2$ price. However, as partial derivatives are involved, we immediately see that the SCC is associated with a \textit{change emanating from the equilibrium state}. It is a marginal price. It is unclear whether this price will cover the cost along the climate change mitigation paths.

	\clearpage	
	\section{An Implied CO$_{\textbf{2}}$-Price that covers the Cost of a Transition to Net Zero}

	\subsection{Definition}

	We calculate the transition cost to net zero and try to assign a uniform CO$_{\text{2}}$-price to it. Here and in the following, we adopt the notation from~\cite{FriesQuanteLong2023}.

	Let $C(t)$ denote the time-$t$ cost, e.g., in a DICE model \cite{Nordhaus2017} we would have $C(t) = C_{\mathrm{A}}(t) + C_{\mathrm{D}}(t)$, where $C_{\mathrm{A}}$ are the cost associated with abatement and $C_{\mathrm{D}}$ are the cost associated with damage. Let $E(t)$ denote the time-$t$ emissions. We value the financial contract that exchanged the (accumulated) costs with a (forward) price $K \cdot N(t)$ paid for the emissions. The value of this financial contract is:
	\begin{equation}
		\label{eq:costVersusEmissionPriceSwap}
		V_{\mathrm{Gap}}(K;0) \ := \ \int_{0}^{T} \left( C(t) - K^{*} \cdot N(t) \cdot E(t) \right) \frac{N(0)}{N(t)} \ \mathrm{d}t \text{.}
	\end{equation}
	Here, $t \mapsto N(t)$ denotes the value process of a risk-free account that serves as numéraire. The factor $\frac{N(0)}{N(t)}$ is the discount factor\footnote{In a model with constant deterministic continuously compounding risk-free interest rate $r$ we have that $N(t) = \exp(r \cdot t)$ describes the value process of a bank account with initial value $N(0) = 1$.}.

	The assumption that the CO$_{2}$-price is given by a constant $K$ times $N(t)$ reflects that the time-value adjusted price is constant.
		
	The time horizon $T$ is chosen such that, approximately, $E(t) = 0, \, C(t) = 0 \ \text{for} \ t \geq T$, such that we distribute \textit{all} cost to \textit{all} contributing emissions. See~Section~\ref{sec:impliedCO2:timeHorizon}.
	
	$V_{\mathrm{Gap}}(K;0)$ can be interpreted as the \textit{gap} that is left by a uniform (time-value adjusted) CO$_{\text{2}}$-price $K$. For $K = 0$ we have that $V_{\mathrm{Gap}}(0;0)$ gives the accumulated cost.
	
	We now seek to find the constant $K_{\mathrm{par}}$ such that $V_{\mathrm{SCC-Gap}}(K_{\mathrm{par}};0) = 0$. In other words, $K_{\mathrm{par}}$ is the (forward) price that equally all generations have to pay for the emissions to finance the costs towards a net-zero transition.
	
	We have
	\begin{equation}
		K_{\mathrm{par}} = \frac{\int_{0}^{T} \frac{C(t)}{N(t)} \ \mathrm{d}t}{\int_{0}^{T} E(t) \ \mathrm{d}t} \text{.}
	\end{equation}
	The price $K_{\mathrm{par}}$ can be interpreted as an \textit{implied CO$_2$-price}. It is implied by the cost $C(t)$ related to climate change and its mitigation. It can be interpreted as the price resulting from a strict application of a \textit{polluter pays principle}.

	\subsection{Choice of the Time-Horizon}
	\label{sec:impliedCO2:timeHorizon}

\begin{figure}[htb]
	    \centering
		\subfloat[\textbf{Cost}, $t \mapsto C(t)$.]{\includegraphics[width=0.45\textwidth]{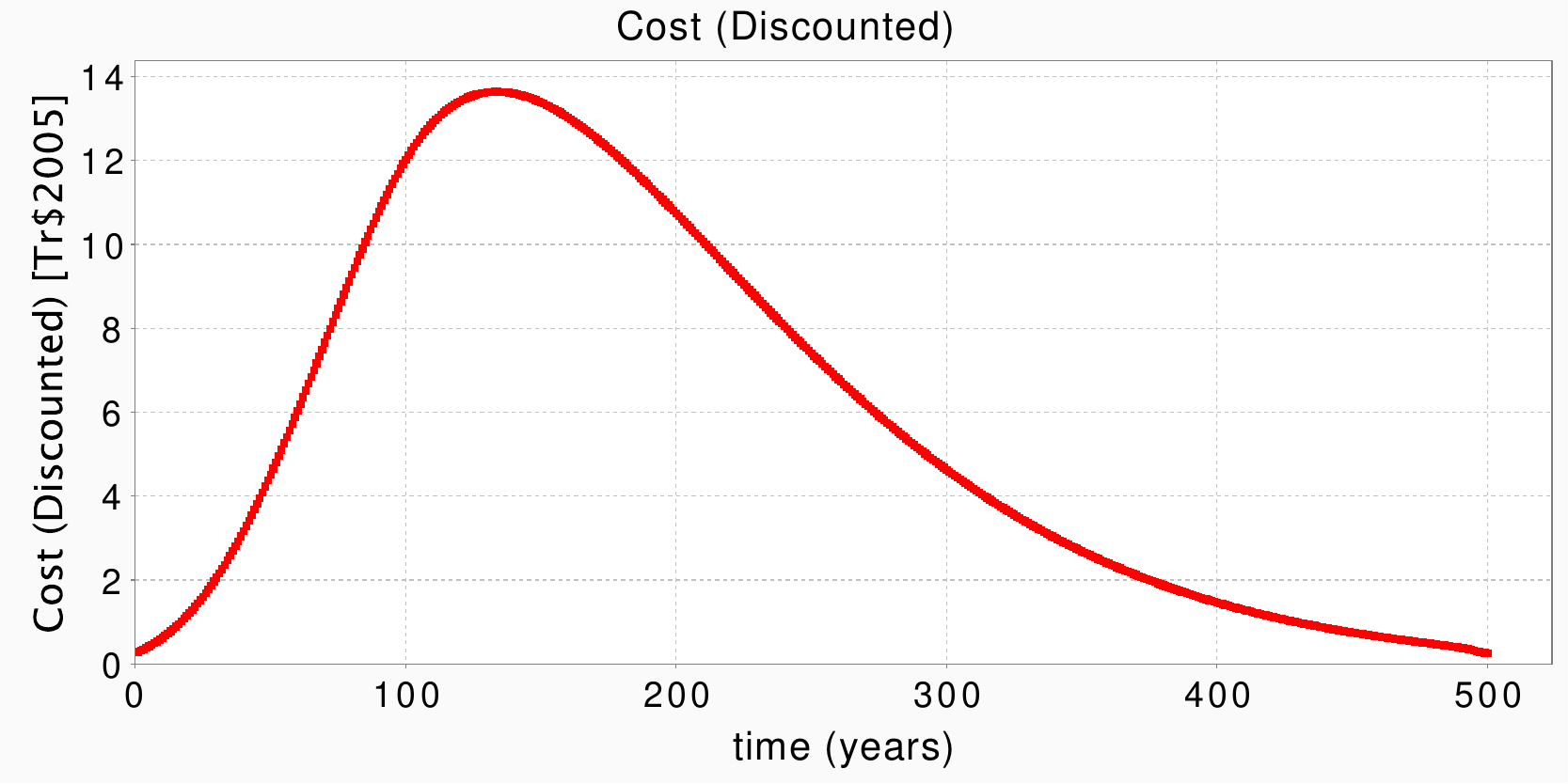}}
		\hfill
		\subfloat[\textbf{Emission}, $t \mapsto E(t)$.]{\includegraphics[width=0.45\textwidth]{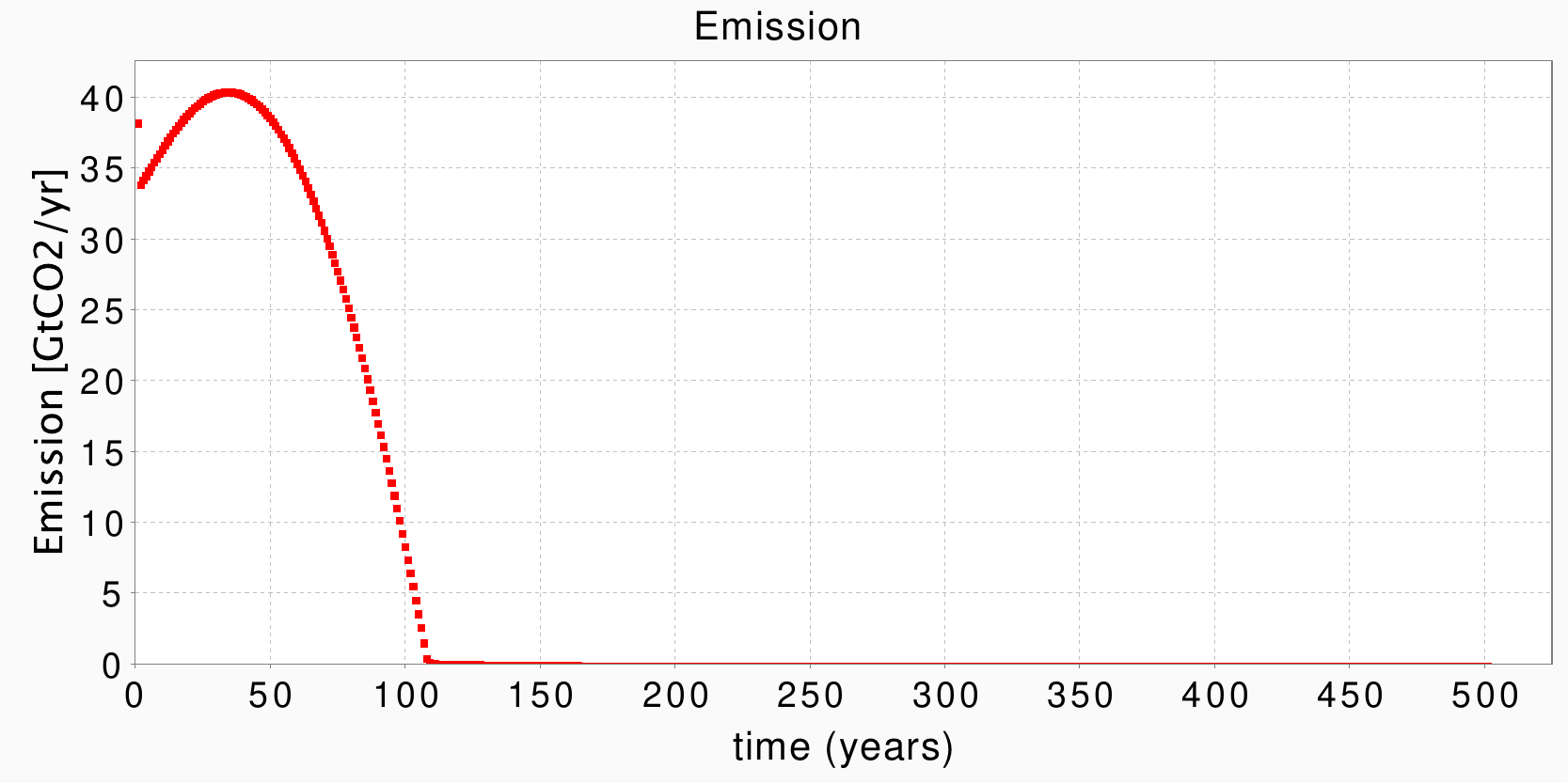}}
		\caption{\textbf{Cost and Emission.} The cost and emission in the calibrated model as a function of time showing that there is a sufficiently fast decay.}
		\label{fig:emissionAndCostOverTime}
	\end{figure}

	The definition of $K_{\mathrm{par}}$  is sensitive to the choice of the time horizon $T$ as long as $E(t) > 0, \, C(t) > 0$. Due to the emissions from land use, $E(t) = 0$ is never fulfilled, but $E(t)$ becomes negligibly small after $T = 500$. Similarly, the discounted cost $\frac{C(t)}{N(t)}$ decline significantly fast, see Figure~\ref{fig:emissionAndCostOverTime}.

\begin{figure}[htb]
	    \centering
		\includegraphics[width=0.8\textwidth]{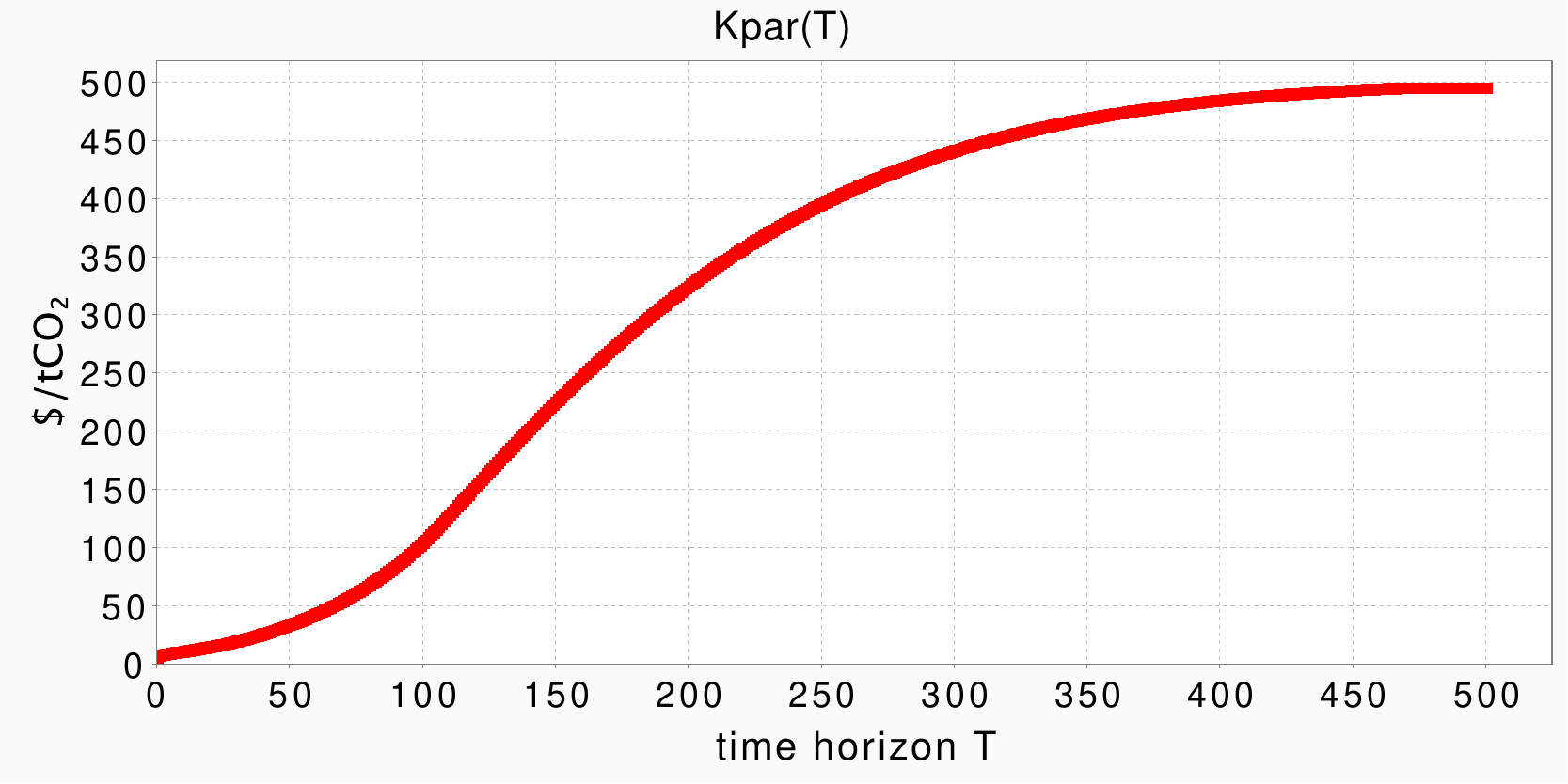}
		\caption{\textbf{Dependency on the Time-Horizon.} The implied CO$_2$-price as a function of the chosen time-horizon $T$. The price convergences as emission and cost go to zero. To read this figure: Requiring that the paying for the emissions of the next 150 years cover the cost of these 150 years would require a CO$_2$ price of $K_{\mathrm{par}}(150) \approx 225$ \$/tCO$_2$.}
		\label{fig:impliedCO2PriceAsFunctionOfTimeHorizon}
	\end{figure}

	If $T$ is significantly large, $K_{\mathrm{par}}$ converges. Its limit is the price obtained by distributing all (future) costs to all (future) emitters. Figure~\ref{fig:impliedCO2PriceAsFunctionOfTimeHorizon} given the price $K_{\mathrm{par}}$ as a function of the chosen time-horizon $T \mapsto K_{\mathrm{par}}(T)$.

	\subsection{Separating Costs from Past Emissions}

	The price $K_{\mathrm{par}}$ also carries the cost of damages resulting from past emissions and distributes them to the future emissions $E(t)$. To separate these costs, we may run the model with an abatement policy given by perfect (100\%) abatement of emissions from the first year on, i.e., $\mu \equiv 1$.

	Let $C_{\mathrm{D}}^{*}(t)$ denote the damage cost observed in this model. We then define
	\begin{equation}
		K_{\mathrm{par}}^{*} \ := \ \frac{\int_{0}^{T} \frac{C(t) - C_{\mathrm{D}}^{*}(t)}{N(t)} \ \mathrm{d}t}{\int_{0}^{T} E(t) \ \mathrm{d}t} \text{.}
	\end{equation}
	The value $K_{\mathrm{par}}^{*}$ can be interpreted as a fair (polluter pays) CO$_2$ price that covers the future cost of abatement and damages, excluding those damages resulting from past emissions.

	We report the values $K_{\mathrm{par}}$  and $K_{\mathrm{par}}^{*}$ for a classical (calibrated) DICE model in Section~\ref{sec:co2pricing:numericalExperiments}.

	\clearpage	
	\subsection{Social Cost of Carbon}

	The Social Cost of Carbon is defined as 
	\begin{equation*}
		SCC(t) := \frac{\partial V(0)}{\partial E(t)} \big/ \frac{\partial V(0)}{\partial Z(t)} \text{,}
	\end{equation*}
	where $V(0)$ is the total discounted welfare, $E(t)$ is the time $t$ emissions measured in tCO$_{\text{2}}$ and $Z(t)$ is the time $t$ consumption measured in Dollar.
	
	The SCC expresses the change in the welfare due to a change of the time $t$ emission per change of time $t$ consumption (GDP), resulting in the equivalent welfare change. 	The result of this definition is a time-$t$ value; that is, the SCC is \textit{undiscounted}.

	In the original work on the dice model one sometimes finds $\frac{\partial V(0)}{\partial E(t)} \big/ \frac{\partial V(0)}{\partial Z(t)} \equiv \frac{\partial Z(t)}{\partial E(t)}$, but this expression is mathematically not correct (it helps however to get an intuitive understanding for the SCC).

	\medskip
	
	If $SCC(t)$ is used as a CO$_{\text{2}}$ price, then the total value paid for CO$_{\text{2}}$ emissions is
	\begin{equation*}
		V_{\mathrm{SCC}}(0) \ := \ \int_{0}^{T} \left( SCC(t) \cdot E(t) \right) \frac{N(0)}{N(t)} \ \mathrm{d}t \text{.}
	\end{equation*}
	
\begin{figure}[htb]
	    \centering
	    \includegraphics[width=\textwidth]{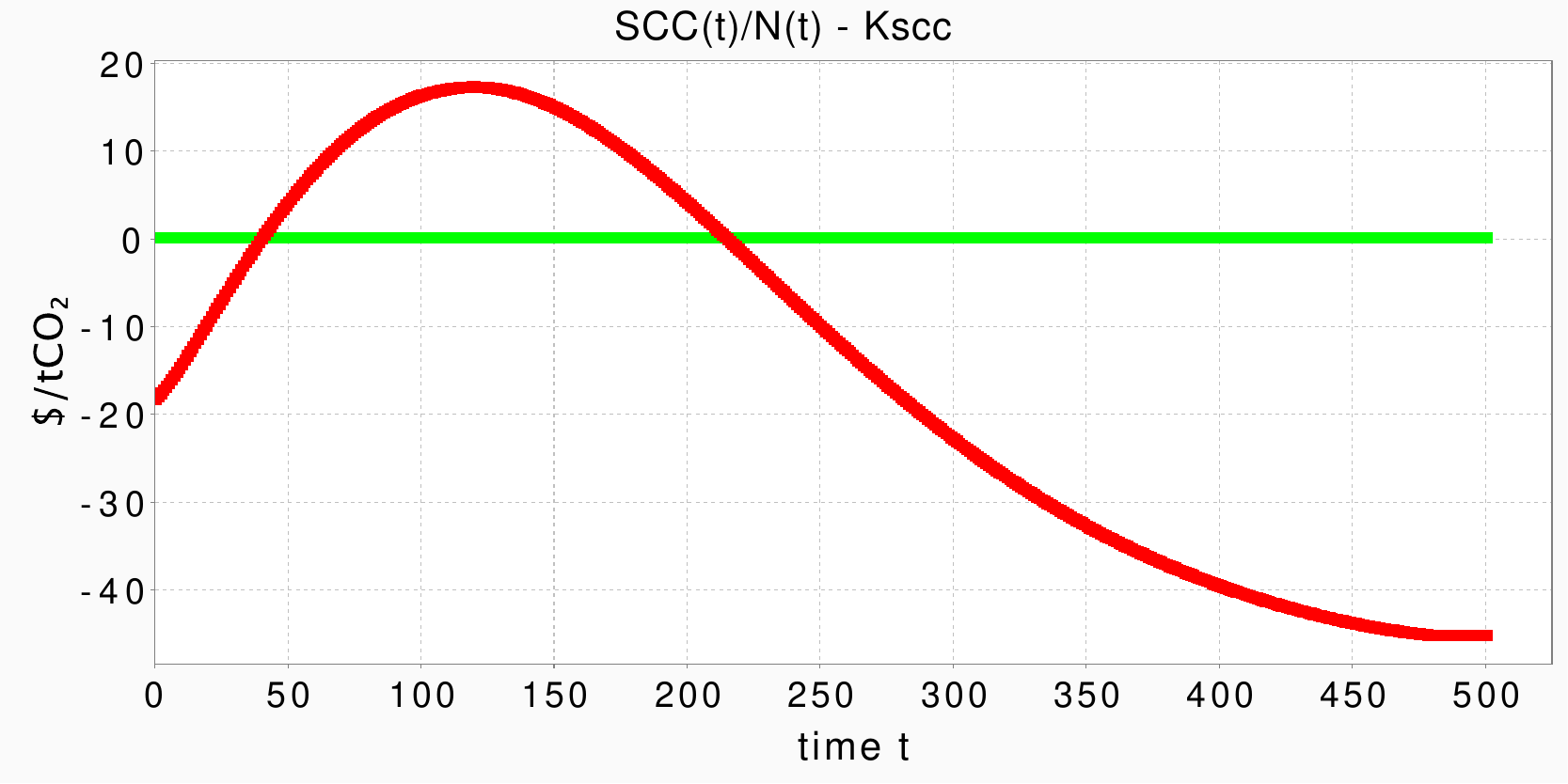}
	    \caption{\textbf{Deviation of the SCC from the corresponding Par-Value.} The difference of the (discounted) social cost of carbon $t \mapsto SCC(t)$ from the corresponding (constant) forward price $K_{\mathrm{SCC}}$. Generations living in years 40 to 220 pay more, while others pay less.}
	    \label{fig:sccVersusKscc}
	\end{figure}

	For this, we can calculate an equivalent swap rate $K_{\mathrm{SCC}}$ such that
	\begin{equation}
		\label{eq:co2pricing:sccParRateEquation}
		\int_{0}^{T} \left( SCC(t) \cdot E(t) - K_{\mathrm{SCC}} \cdot N(t) \cdot E(t) \right) \frac{N(0)}{N(t)} \ \mathrm{d}t \ \stackrel{!}{=} \ 0 \text{.}
	\end{equation}
	The solution of \eqref{eq:co2pricing:sccParRateEquation} is given by $K = K_{\mathrm{SCC}}$ with
	\begin{equation}
		K_{\mathrm{SCC}} \ = \ \frac{\int_{0}^{T} \left( SCC(t) \cdot E(t) \right) \frac{1}{N(t)} \ \mathrm{d}t}{\int_{0}^{T} \ E(t) \ \mathrm{d}t} \text{.}
	\end{equation}
	Here $K_{\mathrm{SCC}}$ represents a fair constant price for all times, while the social cost of carbon varies over time. Figure~\ref{fig:sccVersusKscc} shows the (discounted) difference of the two $SCC(t) / N(t) - K_{\mathrm{SCC}}$.

	\clearpage	
	\subsection{Numerical Experiment: Does the SCC cover the cost?}
	\label{sec:co2pricing:numericalExperiments}

	The social cost of carbon is sometimes considered to represent the CO$_{2}$-price. However, from the definition, it is clear that it is only a marginal price. Hence, it remains open if paying the price $SCC(t)$ for the emissions $E(t)$ covers the cost $C(t)$.
	
	\begin{itemize}
		\item If $K_{\mathrm{SCC}} \geq K_{\mathrm{par}}$, then taking the $SCC(t)$ as the CO$_{\text{2}}$ price will cover the total cost of the mitigation path.

		\item If $K_{\mathrm{SCC}} < K_{\mathrm{par}}$, then taking the $SCC(t)$ as the CO$_{\text{2}}$ price will not cover the total cost of the mitigation path.
	\end{itemize}

	It is maybe obvious that $K_{\mathrm{SCC}}$ does not agree with $K_{\mathrm{par}}$. The mismatch between $K_{\mathrm{SCC}}$ and $K_{\mathrm{par}}$ can be significantly large.
	
	For the DICE-2016 model, we find:\footnote{The results were obtained with a model using $T = 500$, a time discretization step size $\Delta t = 1.0$, and an interest rate of 1.5 \% (annual linear compounding). Results may vary slightly depending on calibration accuracy.}
	\begin{alignat*}{2}
		SCC(2015) & \ \approx \ & 27 & \ \$/\mathrm{tCO}_{2} \text{,} \\
		K_{\mathrm{SCC}} & \ \approx \ & 45 & \ \$/\mathrm{tCO}_{2} \text{,} \\
		K_{\mathrm{par}} & \ \approx \ & 500 & \ \$/\mathrm{tCO}_{2} \text{,} \\
		K_{\mathrm{par}}^{*} & \ \approx \ & 350 & \ \$/\mathrm{tCO}_{2} \text{.}
	\end{alignat*}
	To summarise, taking the Social Cost of Carbon as a CO$_{\text{2}}$ price does not cover the cost of climate change mitigation.
	
	The results of this section can be found in the numerical experiment
	\begin{center}
		\texttt{ClimateModelExperimentCO2Pricing}
	\end{center}
	available at \cite{finmathClimateNonLinear}.

	\clearpage	
	\subsection{Discussion}

	The price $K_{\mathrm{par}}$ in the financial product in \eqref{eq:costVersusEmissionPriceSwap} has an intuitive interpretation. It is distributing the cost associated with the climate mitigation proportional to the CO$_{2}$ emissions. It can be understood as a \textit{polluter pays principle}. Due to the time-value adjustment, it may be considered a fair price.

	The price $K_{\mathrm{SCC}}$ is the corresponding price matching the revenue if emissions have a price that corresponds to the $SCC(t)$.

	That $SCC(0)$ is lower than $K_{\mathrm{SCC}}$ implies that the present generation is paying less (compared to its contribution measured in terms of emissions). The difference
	\begin{equation*}
		K_{\mathrm{SCC}} \ - \ \frac{SCC(t)}{N(t)}
	\end{equation*}
	can be interpreted as a measure of inter-generational inequality. For a discussion on inter-generational inequality, see~\cite{FriesQuante2023}.

	That $K_{\mathrm{SCC}} < K_{\mathrm{par}}$ implies that taking the SCC as a CO$_2$ price does not cover the total cost. The gap is significant. The society covers the remaining cost, but without any relation to the emissions.

	\clearpage	
	\section{An Implied Interest Rate of Carbon}
	\label{sec:co2pricing:interestRateOfCarbon}

	\subsection{Definition}

	In the following, we consider an alternative to the social cost of carbon. Let $\mu$ denote the calibrated abatement policy, $C_{\mathrm{A}}(t)$ the time-$t$ abatement cost and $C_{\mathrm{D}}(t)$ the time-$t$ damage cost.
	A local change $\mathrm{d}\mu(t)$ in the abatement factor $\mu(t)$ generates a change in the emissions, a change to the abatement cost in $t$, $\frac{\mathrm{d} C_{\mathrm{A}}(t)}{\mathrm{d} \mu(t)}$ and a change in the future damage cost in $s \geq t$, $\frac{\mathrm{d}C_{\mathrm{D}}(s)}{\mathrm{d}\mu(t)}$.

	If an increase in the emission in $t$ is linked to a reduction in the abatement level, it saves abatement cost but is repaid with damage cost. This can be considered a loan: increasing emissions, we borrow abatement cost and pay back by damage cost.

	For this loan, we may calculate an \textit{internal rate of return} $r^{\mathrm{SCC}}(t)$ as the solution of the equation
	\begin{equation*}
		\frac{\mathrm{d} C_{\mathrm{A}}(t)}{\mathrm{d} \mu(t)} + \sum_{t_{k} \geq t} \frac{\mathrm{d}C_{\mathrm{D}}(t_{k})}{\mathrm{d}\mu(t)} \exp(- r^{\mathrm{SCC}}(t) \cdot (t_{k}-t)) \ = \ 0 \text{.}
	\end{equation*}
	We may then compare the rate $r^{\mathrm{SCC}}(t)$ with the model's interest rate $r$. If $r^{\mathrm{SCC}}(t) > r$ (and $r$ is the rate to finance a loan), then funding abatement with a loan will be profitable for the society.

	\clearpage
	\subsection{The Implied Interest Rate of Carbon in the Classical DICE Model}

\begin{figure}[htb]
	    \centering
		\includegraphics[width=\textwidth]{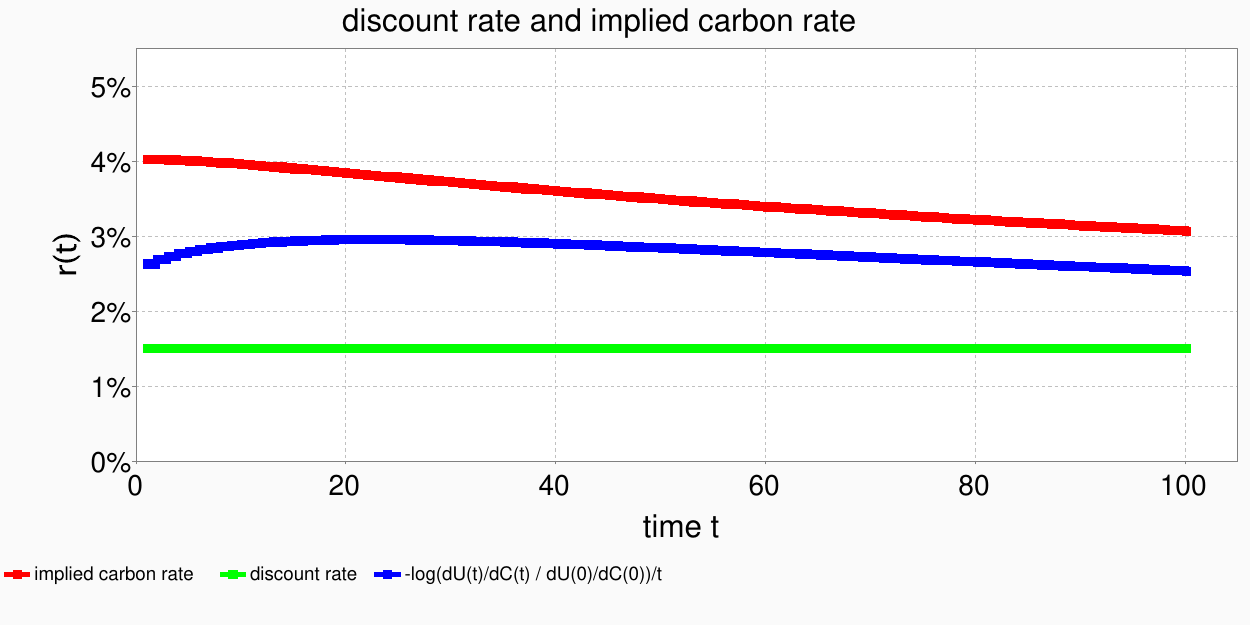}
		\caption{\textbf{Implied Carbon Rate.} The internal rate of return associated with carbon abatement (red) for a calibrated DICE model with a discount rate of 1.5\% (green). A main contribution comes from the model's utility rate (blue).}
		\label{fig:impliedCarbonRate}
	\end{figure}

	We calculate $t \mapsto r^{\mathrm{SCC}}(t) $ in a classical DICE model. Figure~\ref{fig:impliedCarbonRate} shows the result for a model with a calibrated abatement policy. The internal rate of return of carbon is much higher than the model's discount rate. For a DICE-2016 with the discount rate $r = 1.5\%$, we find $r^{\mathrm{SCC}}(t) \approx 4\%$ for the first ten years.

	We derive the rate $r^{\mathrm{SCC}}$ for the DICE model from the model components. Differentiating the model's objective function with respect to $\mu(t_{j})$ we find from $\mathrm{d}V(0) / \mathrm{d}\mu (t_{j}) \ = \ 0$ that
	\begin{align*}
		0 & \ = \ \int_0^T \frac{\mathrm{d}U(t)}{\mathrm{d}\mu (t_{j})} \frac{N(0)}{N(t)} \mathrm{d}t \ = \ \int_{t_{j}}^{T} \frac{\mathrm{d}U(t)}{\mathrm{d}\mu (t_{j})} \frac{N(0)}{N(t)} \mathrm{d}t \\
		&  \ = \ \int_{t_{j}}^{T} \sum_{t_{k} \geq t_{j}} \frac{\mathrm{d}U(t)}{\mathrm{d}C (t_{k})} \frac{\mathrm{d}C (t_{k})}{\mathrm{d}\mu (t_{j})} \frac{N(0)}{N(t)} \mathrm{d}t
		\ = \ \sum_{t_{k} \geq t_{j}} \frac{\mathrm{d}C (t_{k})}{\mathrm{d}\mu (t_{j})} \left( \int_{t_{j}}^{T}  \frac{\mathrm{d}U(t)}{\mathrm{d}C (t_{k})} \frac{N(0)}{N(t)} \mathrm{d}t \right) \text{.}
	\end{align*}
	That is
	\begin{equation}
		\sum_{t_{k} \geq t_{j}} \frac{\mathrm{d}C (t_{k})}{\mathrm{d}\mu (t_{j})} \cdot \exp \left( - r^{\mathrm{SCC}}(t_{j}) (t_{k}-t_{j}) \right)
		\ = \ \sum_{t_{k} \geq t_{j}} \frac{\mathrm{d}C (t_{k})}{\mathrm{d}\mu (t_{j})} \cdot \left( \frac{R(t_{j},t_{k})}{R(t_{j},t_{j})} \right) \text{,}
	\end{equation}
	with
	\begin{equation*}
		R(t_{j},t_{k}) \ = \ \int_{t_{j}}^{T}  \frac{\mathrm{d}U(t)}{\mathrm{d}C (t_{k})} \frac{N(0)}{N(t)} \mathrm{d}t \text{.}
	\end{equation*}
	Here $U$ denotes the model's utility function. The rate $r^{\mathrm{SCC}}$ is the sum of the discount rate and a weighted average of the utility rate.\footnote{Note that $r^{\mathrm{SCC}}(t_{j}) \ \neq \ - \frac{1}{t_{k}-t_{j}} \log \left( \frac{R(t_{j},t_{k})}{R(t_{j},t_{j})} \right)$.}

	\medskip

	The results of this section can be found in the numerical experiment
	\begin{center}
		\texttt{ClimateModelExperimentSCCRate}
	\end{center}
	available at \cite{finmathClimateNonLinear}.

	\clearpage
	\subsection{Discussion}

	The rate $r^{\mathrm{SCC}}$ can indicate if it is profitable to finance abatement by a loan. If borrowing at a lower rate $r < r^{\mathrm{SCC}}$ is possible, then abatement should be financed by a loan,~\cite{FriesQuante2023}.
~
	For the classical DICE model, we found that the rate $r^{\mathrm{SCC}}$ is much higher than the discount rate. This finding does not contradict the fact that the model is in its calibrated equilibrium. It is explained by the objective function, balancing discounted marginal utility changes, but not cost, see~\cite{FriesQuanteLong2023}.	


	\clearpage
	\printbibliography

\newpage

\section*{Notes}
%
\subsection*{Suggested Citation}

\begin{itemize}
	\item[] \sloppypar \textsc{Fries, Christian P.}: Implied CO$_{\textbf{2}}$-Price and Interest Rate of Carbon. (December, 2023).
	\url{https://ssrn.com/abstract=3284722}
	\newline
	\url{http://www.christian-fries.de/finmath}
\end{itemize}
%
%
\begin{small}
%
%
%
%

	
	
	\noindent Keywords:
	Integrated Assessment Models,
	\\ \phantom{Keywords:\ }
	CO$_2$-Price
	\\ \phantom{Keywords:\ }
	Social Cost of Carbon,
	\\ \phantom{Keywords:\ }
	Interest Rate of Carbon,
	\\ \phantom{Keywords:\ }
	Intergenerational Equity

	\noindent Classification:
	MSC 91-10
	
	
	
	\vfill
	
	\bigskip
	\centerline{\small\thepage~pages. \totalfigures~figures. \totaltables~tables.}
	
\end{small}

\end{document}